\documentclass[preprint,aps,prl,floatfix,showpacs]{revtex4}
\usepackage{graphicx}

\begin{document}

\title{
An anomalous band inversion protected by symmetry in a topological insulator of 
the Kane-Mele model
}

\author{Jie-Xiang Yu}
\author{J. G. Che}\altaffiliation{Corresponding author.
E-mail: jgche@fudan.edu.cn.}
\affiliation{Surface Physics Laboratory (National Key Laboratory),
Key Laboratory of Computational Physical Sciences (MOE),
Department of Physics and
Collaborative Innovation Center of Advanced Microstructures, Fudan University,
Shanghai 200433, People's Republic of China}

\pacs{73.43.-f,71.70.Ej,71.15.Mb,73.22.Pr}

\begin{abstract}
Depositing Au on a graphene derivative, which involves substituting four C atoms with three N atoms in 
a $3\times 3$ cell graphene, 
we realized a topological insulator of the Kane-Mele model with a gap of 50~meV surrounding 
the Dirac point of graphene. In this material, we observed an anomalous band inversion (BI) 
protected by the symmetry with character $e$ of group C$_{\rm 3V}$. 
The symmetry constrains two $e$ bands with mirror-symmetry combination (MSC) 
and mirror-antisymmetry combination (MAC) of Au and N orbitals degenerate at $\Gamma$, whereas the interaction 
of $\pi^*$ of graphene on the $e$-MAC band tends to lift this degenerate, resulting in that 
the $\pi^*$ and $e$-MAC band exchange their orbital components near $\Gamma$, causing thus 
a discontinued BI.
\end{abstract}

\maketitle

Kane and Mele suggested in their primary work that a two-dimensional (2D) topological 
insulator (TI) could be realized in graphene.\cite{Kan05} The 2D TI is called a Quantum spin Hall 
(QSH) insulator.\cite{Kan05} However, it is difficult to experimentally observe its topological gap 
because of graphene's very weak spin-orbit coupling (SOC).\cite{Hue06,Min06,Yao07,Gmi09} 
Due to its potential applications, Kane and Mele's suggestion has promoted significant research efforts 
to realize 2D TI in graphene.\cite{Moo10,Wee11,Hu12,You14} Young {\it et al.}~\cite{You14} confirmed that 
the QSH state could be experimentally observed in a charge-neutral graphene under a very large 
magnetic field. A graphene having certain heavy adatoms was predicted by Week {\it et 
al.}~\cite{Wee11} to realize a TI with substantial band gap. A similar method to obtain a gap 
exceeding 0.2~eV without tuning the Fermi level to enter the TI state was suggested by Hu {\it et 
al.}~\cite{Hu12} 
It was reported that the graphene related heterostructure (graphene sandwiched between 
two strong SOC materials such as Bi$_2$Se$_3$) could enhance the intrinsic SOC of graphene and 
open a topological nontrivial gap.\cite{Kou13,Kou14,Ko14}
As a material for use in next-generation electronic devices, graphene is stable and shows 
remarkable properties.  
Realizing a graphene-templated TI in combination with advanced technology have a wide array of 
applications, from spintronics to quantum information process.\cite{Gei07,Xu13} 

In the present work, we propose a platform based on a graphene derivative to realize 
a topological insulator of the Kane-Mele model.
The graphene template involves substituting four C atoms with three N atoms in 
a $3\times 3$ graphene cell ($g$-C$_{14}$N$_3$). 
After Au 
is deposited 
on $g$-C$_{14}$N$_{3}$, 
two bands of the interaction between Au and N are characterized by 
irreducible representation $e$ of group C$_{\rm 3v}$. 
The symmetry constrains the two $e$ bands with mirror-symmetry combination (MSC) and mirror-antisymmetry 
combination (MAC) of Au and N orbitals degenerate at $\Gamma$, whereas the interaction of $\pi^*$ on the $e$-MAC band tends 
to lift this degenerate, 
resulting in that a piece of the $e$-MAC band near $\Gamma$ exchanges with a piece of the $\pi^*$ band, 
causing thus a discontinued BI between the $\pi^*$ and e-MAC band. 
This anomalous BI protected by symmetry leads to a topological nontrivial gap of 50~meV 
around the Dirac point of graphene. 

The results were obtained by our first-principles calculations on the framework of density 
functional theory with the projector augmented plane-wave (PAW) potential~\cite{PAW} as 
implemented in VASP package\cite{VASP}. The exchange-correlation energy was described by 
local density approximation (LDA),\cite{LDA} which treated the atomic positions well in the 
graphene-based structures.\cite{Wang09} The wave functions were expanded in a plane-wave 
with an energy cutoff of 600~eV throughout calculations. The $k$ points in surface Brillouin zone 
(BZ) of the $3\times 3$ unit cell of graphene-based structures were sampled on a $\Gamma$-centered 
$7\times 7$ mesh. All atoms could be relaxed until the Hellmann-Feynman forces on the atoms 
were less than 0.001~eV/\AA. 
We performed the maximally localized Wannier function (MLWF) 
process implemented in Wannier90 package for fitting tight-binding Hamiltonian and analyzing 
the evolution of the Wannier function center (WFC).\cite{EoWFC} 

\begin{figure}[bt]
\centerline{\includegraphics[scale=1.20,angle=0]{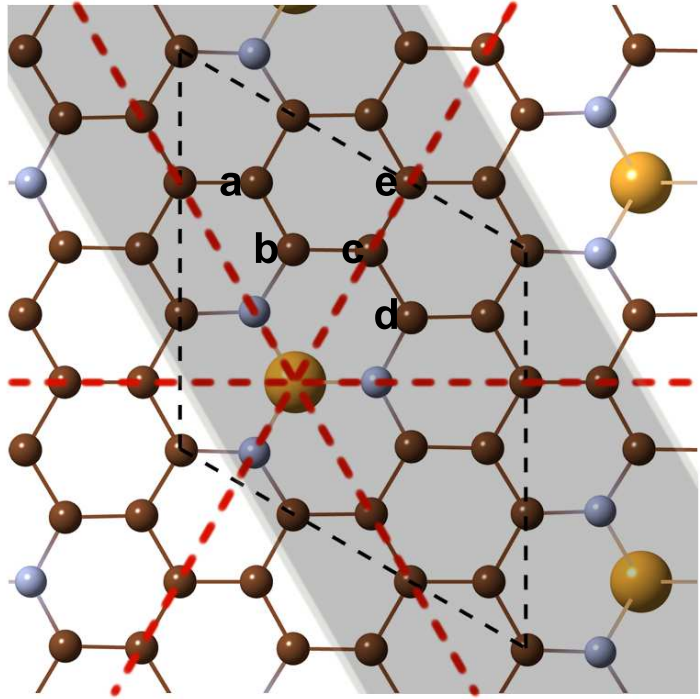}}
\caption{(Color online)
Topview of the atomic configuration for $g$-C$_{14}$N$_3$. Brown, blue 
and yellow balls represent C, N and Au, respectively. Thin-dashed lines denote the boundary of a 
$3\times 3$ unit cell of graphene with three mirror-symmetric planes, which are identified by 
three red thick-dashed lines. The shadow region indicates the structure of a 1D ribbon with 
an armchair edge. Labels a, b, c, d and e indicate different vacancy-chainis, see text. 
}
\label{config}
\end{figure}

Lee {\it et al.} synthesized a graphitic carbon nitride material, which can be seen as 
four C atoms are substituted by three N atoms in a $2\times 2$ unit cell of graphene, 
$g$-C$_4$N$_3$.\cite{Lee10} 
However, by observing its electronic structures, we found that this material was not suitable for 
inverting the bands around the Fermi level because of its large 2~eV gap.\cite{Du12} 
This large gap was caused by the substitution of four C atoms with three N atoms, which completely breaks 
$\pi$-bonds in a $2\times 2$ graphene cell that is critical for forming the Dirac cone.  
Since the large gap can be seen because very few C($pz$) pairs existed in $g$-C$_4$N$_3$, 
extending its reassembly is worthwhile. 
To realize 2D TI based on graphene with the Dirac-like point near the Fermi level, as suggested by 
Kane and Mele,\cite{Kan05} 
a simple and effective approach is to extend the graphene unit cell from $2\times 2$ to 
$3\times 3$ size ($g$-$3\times 3$),
as shown in Fig.~\ref{config}, hereafter referred to as $g$-C$_{14}$N$_3$. 
$g$-C$_{14}$N$_3$ can be seen as a patterned graphene. As Kim {\it et al.} 
suggested, nanoscale patterns can be made in a horizontally prepatterned graphene using standard 
lithography techniques.\cite{Kim09} 
In that case, the Kane-Mele's model TI may be realized by depositing strong SOC elements 
such as Au on $g$-C$_{14}$N$_3$. The Dirac point appears again because more $\pi$-bonds 
existed in $g$-C$_{14}$N$_3$ than in $g$-C$_{4}$N$_3$. We can see below that the +1e hole with the Dirac point 
0.1 eV above the Fermi level in $g$-C$_{14}$N$_3$ plays an important role in forming TI 
after Au is adsorbed on $g$-C$_{14}$N$_3$.
Considering the stability of $g$-C$_4$N$_3$ obtained by Du {\it et al.},\cite{Du12} 
$g$-C$_{14}$N$_3$ can be also expected to be stable.

It is well known that a stable crystal with vacancies should be those with the least broken 
bonds.\cite{DanglingBonds} We accordingly considered the stability of $g$-C$_{14}$N$_3$.
Following this criterion, only three possible configurations 
by removing four C atoms at sites a-b-c-d (chair), a-b-c-e (boat) and b-c-d-e (star type) 
in Figure 1 need to be taken into account. The criterion also dominates the most stable 
configuration for depositing three N atoms on the graphene with vacancies. That is, 
in order to reduce the number of broken bonds, all three N atoms favor to occupy the vacancy sites. 
Due to the symmetry, there are two inequivalent configurations for three N atoms on four vacancies 
of the boat and chair type. 
We optimized the five configurations. The total energy of the optimized configuration of N 
on the star-type vacancies ($g$-C$_{14}$N$_3$) is at least 2~eV lower than that of N on 
the boat- and chair-type vacancies.

Owing to its importance, we first focus on the electronic structure of $g$-C$_{14}$N$_3$. 
As mentioned above, $g$-C$_{14}$N$_3$ can be seen a derivative of a $3\times 3$ unit cell of 
graphene ($g$-$3\times 3$), in which four C atoms are substituted by three N atoms. N replaces the 
position of C with only a relatively small displacement: the length of the N-C bond is 1.34~\AA, 
whereas that of the C-C bond in graphene 1.42~\AA.\cite{Wal47,Slo58} The displacement of N in 
the direction perpendicular to the graphene plane is $0.08\times 10^{-4}$~\AA, smaller than our 
calculated accuracy. In other words, three N atoms lie in the same plane as that of C. 

\begin{figure}[bt]
\centerline{\includegraphics[scale=1.20,angle=0]{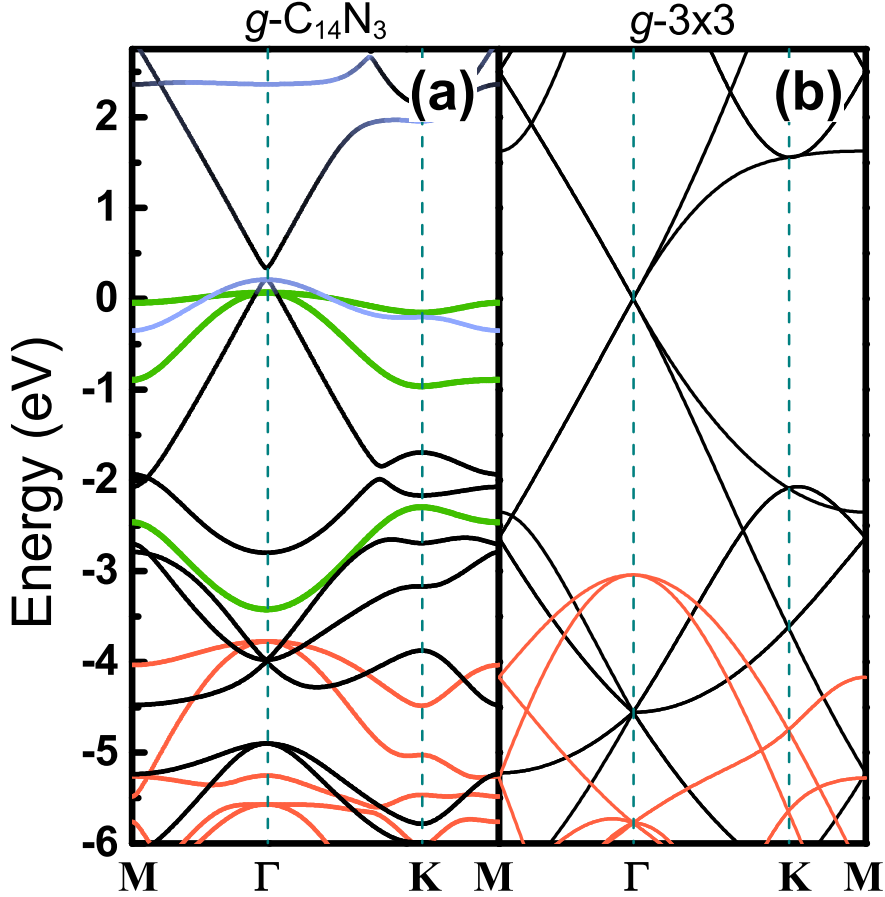}}
\caption{(Color online)
Band structure of $g$-C$_{14}$N$_3$ (a) and g$3\times 3$ (b). Green lines 
denote the bands consisting of N's lone pairs, blue lines denote the bands consisting of
about 25\% N($p_z$) and 27\% C($p_z$) orbitals, 
and red lines denote $\sigma$ bonds of C-C with $sp^2$ hybridization. 
}
\label{C14N3}
\end{figure}

We plotted the band structures of $g$-C$_{14}$N$_3$ in Fig.~\ref{C14N3} (a). The 
band structures of the graphene in the $3\times 3$ size unit cell ($g$-$3\times 3$) are also given in 
Fig.~\ref{C14N3} (b) 
for comparison. Compared with the band structures of the graphene in the $1\times 1$ unit cell 
($g$-$1\times 1$), two features for $g$-$3\times 3$ should be addressed: (1) The Dirac point of 
$g$-$1\times 1$ is folded from the K point of $g$-$1\times 1$ to the $\Gamma$ point of $g$-$3\times 3$; 
and (2) along the $\Gamma$-M axis, the bands surrounding the Fermi level are twofold degenerate, whereas 
along the $\Gamma$-K axis, this degenerate is lifted. This can be expected, since the axis along $\Gamma$-K in 
the BZ of $g$-$3\times 3$ has two unequivalent ways to fold from the axes along K-$\Gamma$ and K-M 
in the BZ of $g$-$1\times 1$. Below, we name them $\pi$- and $\pi$'-bonding states, 
respectively, whereas $\pi$* and $\pi$'* are their antibonding states.

Although four C atoms are replaced by three N atoms in $g$-$3\times 3$, 
the $\pi$  and $\pi$* band are almost 
unaffected by N in $g$-C$_{14}$N$_3$ compared with those in $g$-$3\times 3$. However, a 0.1~eV gap
is opened around the Dirac point and, the Fermi level shifts down by 0.2~eV with respect 
to the top of the occupied $\pi$ band, leaving a +1$e$ hole. The $+1e$ hole can be expected since 
four C atoms with 16 valence electrons are substituted by three N atoms with only 15 valence 
electrons. 

In contrast, the $\pi$'  and $\pi$'* band in $g$-$3\times 3$ disappear in $g$-C$_{14}$N$_3$. Instead, 
there is a bonding band near the Fermi level, consisting of about 25\% N($p_z$) and 75\% 
C($p_z$) orbitals, which also degenerate with $\pi$ at the $\Gamma$ point, and an antibonding band 
2.5~eV above the Fermi level. Both the bonding and antibonding band are shown by blue lines in 
Fig.~\ref{C14N3} (a). They are still referred to as $\pi$' and $\pi$'* in the following discussion.
The separation in energy between the $\pi$'  and $\pi$'* band is 
about 2.5~eV, consistent with the gap in $g$-C$_4$N$_3$,\cite{Du12} which can also be seen as 
a graphene derivative with the same substitution formula. In $g$-C$_4$N$_3$, the Dirac point 
disappeared.\cite{Du12} However, in $g$-C$_{14}$N$_3$, a 0.1~eV gap is opened around
the Dirac point. 

The band difference between $g$-C$_{14}$N$_3$ and $g$-$3\times 3$ is observed mainly in the 
energy region between $-4$ and 2~eV, as shown in the figure by the blue and green lines. Three 
green bands can be attributed to the interaction between three dangling bonds of N's $sp^2$ 
hybridization, the so-called lone pairs. The three dangling bonds of N interact with each other and 
form one singlet $a_1$ and two doublet $e$ states characterized by group C$_{\rm 3V}$. At the $\Gamma$ point, 
$a_1$ and $e$ lie at $-3.5$~eV and near the Fermi level, respectively. The $e$ bands arising 
from the orbital mirror-symmetric combination (MSC) and mirror-antisymmetric combination 
(MAC) of N's $sp^2$ hybrids are degenerate at the $\Gamma$ point and disperse away from it, with the 
MSC band always below the MAC band, as shown in the figure. 
The topological nontrivial gap of graphene calculated by us is less than 10$^{-3}$~meV, 
in agreement with the earlier calculation.\cite{Yao07} However, from the aspect of experiment, 
the gap is too small to be observed.\cite{Hue06,Min06,Yao07,Gmi09} 
We will focus on the SOC effect in terms of introducing 
elements such as Au with strong SOC. 
We will see below that the 
evolution of these bands upon interaction with deposited Au shows that $g$-C$_{14}$N$_3$ is 
a good platform for realizing the Kane-Mele's TI.

Among all inequivalent adsorption (six top, six bridge and three hollow) sites considered
for Au adsorption on $g$-C$_{14}$N$_3$, 
the most stable one, at least 0.8~eV lower than the others, 
is shown in Fig.~\ref{config}: 
Au is lifted by 1.68~\AA\ away fromn the 
graphene layer, and the upward shift of three N atoms to the graphene layer is less than 0.07~\AA. 
Each N is bonded to two neighbor C atoms with a length of C-N of 1.36~\AA. The C-N-C 
angle is $123^o$, which is a slightly larger than that of C atoms in graphene ($120^o$). 
The adsorption of Au on $g$-C$_{14}$N$_3$ gains an energy of 2.6~eV. 

In order to further confirm the stability of Au/$g$-C$_{14}$N$_3$, we performed a phonon calculation implemented 
in VASP package.\cite{VASP} In the calculation, the density functional perturbation 
theory~\cite{Wan11} was used to calculate force constants, of which the concerned 
derivatives were determined by linear response with respect to changes in the ionic positions. 
There are no imaginary frequencies associated with distortion of Au/$g$-C$_{14}$N$_3$,
indicating its stability.

\begin{figure}[bt]
\centerline{\includegraphics[scale=1.20,angle=0]{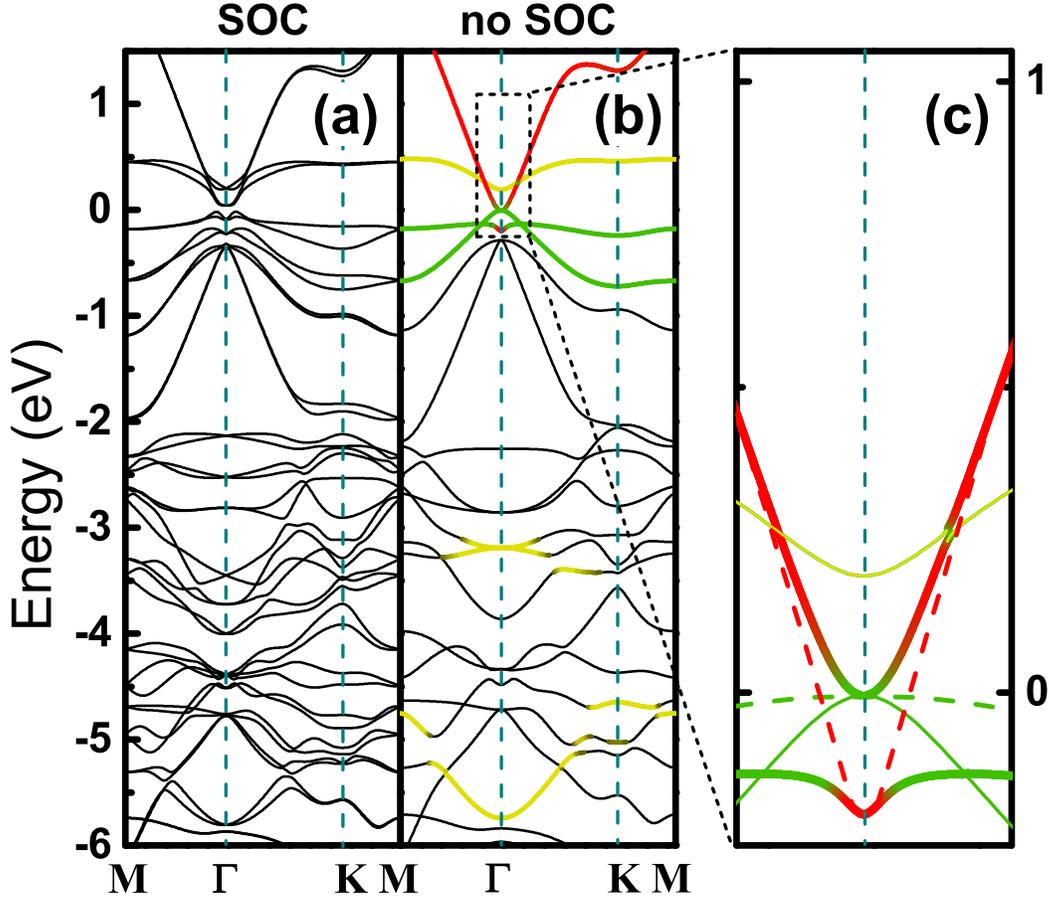}}
\caption{(Color online)
Band structure of Au/$g$-C$_{14}$N$_3$ (a) with and (b) without SOC. The 
green and yellow lines denote the bands primarily consisting of Au and N orbitals, and the red lines
denote the $\pi$* band. Panel (c) is an enlargement of the dashed-rectangle portion in panel (b) in order 
to more clearly illustrate the discontinued thick green and red bands. A conventional band inversion is 
schematically described by the dashed green and red lines in panel (c). 
}
\label{Au}
\end{figure}

Band structures of Au/$g$-C$_{14}$N$_3$ with and without SOC were calculated and are shown 
in Fig.~\ref{Au} (a) and (b), respectively. Fig.~\ref{Au} (c) is an enlargement of 
the dashed-rectangle portion in Fig.~\ref{Au} (b). The band structures with SOC 
(Fig.~\ref{Au}(a)) show a gap of 50~meV surrounding the Dirac point.

Here the key is that the electronic properties of $g$-C$_{14}$N$_3$ showed a $+1e$ hole with the Dirac point 
0.1~eV above the Fermi level. After adsorption of Au in a $d^{10}s^1$ configuration on $g$-C$_{14}$N$_3$, 
the Fermi level returns to the Dirac point. Simultaneously, the $s$ and $d_{xz}$/$d_{yz}$ orbitals 
of Au have to form an $sd^2$ hybridization to adapt to the $g$-C$_{14}$N$_3$ with the threefold 
coordination of N. Thus, $d$-electrons of Au with strong SOC appear around the Fermi level, 
resulting in the topological nontrivial gap.\cite{Bia12,Bia15}

\begin{figure}[bt]
\centerline{\includegraphics[scale=1.20,angle=0]{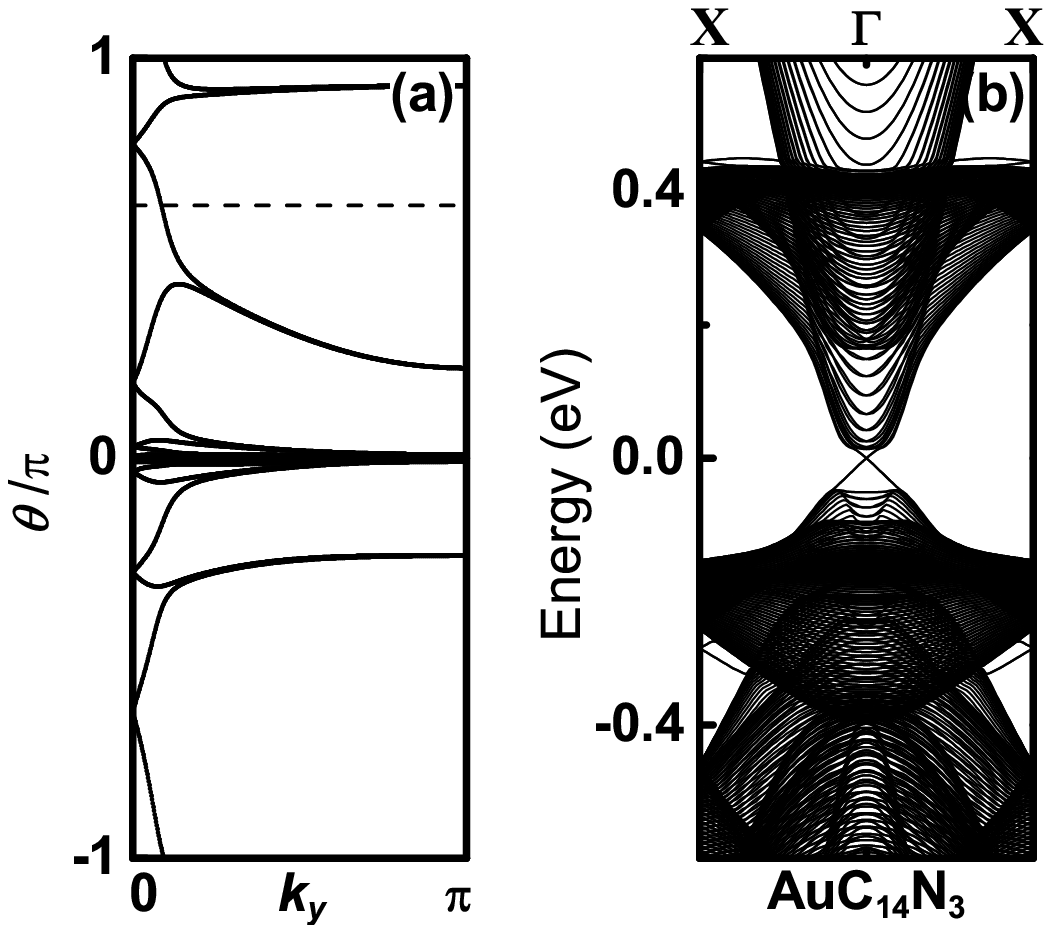}}
\caption{
Evolution of the Wannier function center (a) and band structures of the stack ribbon of 
Au/$g$-C$_{14}$N$_3$ (b). The dashed line in panel (a) is used as a reference. 
}
\label{edge}
\end{figure}

The topological nontrivial properties of Au/$g$-C$_{14}$N$_3$ can be verified by 
the evolution of the Wannier function center (WFC) and the band structures of a 2D stack ribbon 
of Au/$g$-C$_{14}$N$_3$, as shown in Fig.~\ref{edge} (a) and (b), respectively. 
The WFC in reciprocal space during a "time-reversal pumping" 
process was calculated by the $U(2N)$ non-Abelian Berry connection.\cite{EoWFC} 
Au/$g$-C$_{14}$N$_3$ can be seen as a 1D ribbon with an armchair edge periodically stacking 
along the direction perpendicular to the edge of the 1D-ribbon, whose unit cell has two Au, six N 
and 28 C atoms as shown by the shadow region in Fig.~\ref{config}. 
The band structures shown in Fig.~\ref{edge} (b) were 
obtained by calculating a stack ribbon with 50 1D ribbons of Au/$g$-C$_{14}$N$_3$, whose 
interaction parameters were obtained by the MLWF method.\cite{MLWF2,Wei10} 
We confirmed that the stack ribbon with 50 1D-ribbons is wide enough to avoid the interaction 
between the two edges over the stack ribbon. 
The odd winding number of the WFC and the gapless edge states of the stack ribbon,
shown in Fig.~\ref{edge} (a) and (b), respectively,
confirm~\cite{EoWFC} that a topological nontrivial phase appears after Au is deposited on 
$g$-C$_{14}$N$_3$.

Next, we focus on the bands near the Fermi level without SOC (Fig.~\ref{Au} (b)). 
With the help of the orbital 
information (green: Au+N orbitals $>$ 74\%, red: $\pi$* orbitals $>$ 74\%), 
the bands near the Fermi level of Au/$g$-C$_{14}$N$_3$ can be understood. 
At first glance, both the dispersion and orbitals of two bands, which are distributed in the energy 
region between $-0.2$ and $-2.2$~eV and degenerated at the $\Gamma$ point, are similar to that of the 
$\pi$ and $\pi$' bands of $g$-C$_{14}$N$_3$. That is, the two bands are less affected by the 
adsorption of Au compared with that in $g$-C$_{14}$N$_3$. However, owing to the Au 
deposition, the Fermi level is now found to be at the bottom of the $\pi$* band,
indicating that the two bands are now fully filled and that the top of both bands, which 
degenerate at $\Gamma$, lies $-0.2$~eV below the Fermi level. This implies that the +1$e$ hole has been 
fully filled. The topmost occupied bands are now the bands with the $e$ symmetry of C$_{\rm 3V}$, 
indicated by thick and thin green lines. Thus, in comparison with the band order in 
$g$-C$_{14}$N$_3$, a BI between the $\pi$ ($\pi$') and the $e$ bands occurs because of Au 
adsorption. 

Furthermore, the $a_1$ and $e$ bands arising from the interaction of three N's lone pairs in the 
$g$-C$_{14}$N$_3$, as shown by the green lines ranging from 0.0 to $-3.5$~eV in Fig.~\ref{C14N3} 
(a), have disappeared. Instead, two green and three yellow bands exist below the Fermi level; 
these five bands can be traced to the interactions between Au and N, implying that Au has broken 
the lone pair bonds of the three N atoms. 

In order to adapt to the threefold coordination of the N atoms, 
the $s$ and $d_{xz}$/$d_{yx}$ orbitals of Au have to form an $sd^2$ hybridization.
Therefore, the two interaction bands between Au and N, 
the thick and thin green bands near the Fermi level in Fig.~\ref{Au} (c), 
keep the same $e$ symmetry as the lone pair bands of N in 
$g$-C$_{14}$N$_3$, i.e., the bands with MSC and MAC of Au and N orbitals, 
corresponding to the $e$ irreducible representation of C$_{\rm 3V}$.\cite{Group}. 

The orbital analysis of the $e$ band (thick green) and $\pi$* band (thick red) band, as shown in 
Fig.~\ref{Au} (b) and (c), identifies from their color that their orbital components are exchanged 
at 0.0~eV and $-0.2$~eV, respectively. 
This means that without SOC the two bands (thick green and 
red) are inverted near the $\Gamma$ point. In contrast to a conventional BI, however, both inverted 
bands are {\it discontinuous} in the orbital components (indicated by color). 
Considering the $e$ symmetry for the thick (MAC) and thin green (MSC) bands, and the $a_2$
symmetry for the thick red (MAC) band, the anomalous BI can be understood. 

In $g$-C$_{14}$N$_3$, the $e$ bands consist of MSC and MAC of N's orbitals. 
In the entire BZ, the $e$ band with MSC is lower in energy than that with MAC, as 
shown in Fig.~\ref{C14N3} (a). After Au adsorption, the bonds between the three N's lone pairs 
are broken, forming the interaction bonds between Au and N consisting of the $s$ and 
$d_{xz}$/$d_{yz}$ orbitals of Au ($sd^2$ hybrids) and the $s$ and $p_x$/$p_y$ orbitals of N 
($sp^2$ hybrids).  Hereafter, the two bands are referred to as Au+N. 
Most importantly, the Au+N bands have the same $e$ symmetry (MSC and MAC) as the bands 
of the three N's lone pairs in $g$-C$_{14}$N$_3$, as discussed above. 
Therefore, if there are no perturbations of $\pi$*, the 
dispersion of the $e$ band with MAC should be schematically described by the 
dashed green line in Fig.~\ref{Au} (c), i.e., the dashed green line and the thin green line are 
degenerate at the $\Gamma$ point and disperse away from it, similar to that in $g$-C$_{14}$N$_3$, as 
shown by the two green lines near the Fermi level in Fig.~\ref{C14N3} (a). 

As we have seen from the band structures of $g$-C$_{14}$N$_3$ in Fig.~\ref{C14N3} (a), the $e$ 
band with MSC is lower in energy than those with MAC along the axis M-$\Gamma$-K. 
In contrast, for 
Au/$g$-C$_{14}$N$_3$, approaching the $\Gamma$ point the $e$ band with MAC (thick green line) 
first crosses over and then is lower than the $e$ band with MSC (thin green line), 
as shown in Fig.~\ref{Au} (c). We now try to understand the physics behind the observation. 

Since Au does not lie in the same plane as N and C (1.68~\AA\ above), the thick and thin 
green bands are mixed with a few $p_z$ orbitals of both N and C, indicating the interaction 
between Au+N and $\pi$*. 
Thus, the dispersion of the Au+N band with MAC is affected.
The dashed green band in Fig.~\ref{Au} (c) can be viewed as an unaffected $e$ band with MAC,
keeping its energy order relative to the thin green line (the $e$ band with MSC), 
similar to the two $e$ bands in $g$-C$_{14}$N$_3$ shown in Fig.~\ref{C14N3} (a).
Due to the interaction with $\pi$*, the thick green band is lower in energy than the dashed green band. 
This can be seen as the first interaction on the $e$ band with MAC (thick green line). 
In fact, there is also an interaction between $\pi$* and the $e$ band with MSC 
(thin green line). However, since $\pi$* consists of the orbitals of MAC, the interaction bteween
the bands with MSC and MAC cancelled out each other. 

It is well known that for the bands with $e$ symmetry of C$_{\rm 3V}$, its MSC and MAC 
states are constrained to degenerate at the $\Gamma$ point.\cite{Group} This symmetry constraint 
can be seen as the second interaction with an opposite direction to the first one.
This symmetry constraint reaches its maximum at the $\Gamma$ point and weakens away from it. 

Thus, we can understand that the anomalous BI is a result of the competition between the 
first and second interaction mentioned above. 
On approaching the $\Gamma$ point, due to this competition 
a piece of the thick green band with MAC (Au+N) switches 
from the band at $-0.2$~eV to the band at 0.0~eV, being degenerate with the thin green 
band with MSC because of the symmetric constraint, and a piece of the thick red 
band with MSC ($\pi$*) does the opposite in order to retain the band continuity 
at $-0.2$~eV, as shown in Fig.~\ref{Au} (c). 
That is, the constraint of symmetry and band continuity makes
an exchange of the orbital components between the thick green 
and red band near the $\Gamma$ point, leading to a BI.
In contrast to a conventional BI, both the thick green and red band  
are discontinuous, and the discontinued BI is protected by symmetry. 
To the best of our knowledge, this anomalous BI has not been reported before.

The intrinsic SOC of the graphene sandwiched between Bi$_2$S$_3$ layers with strong 
SOC could be enhanced, leading to a BI between the conducting band of Bi$_2$S$_3$
and the valence band of graphene.\cite{Kou13} The mechanism of the graphene 
related heterostructures is similar.\cite{Kou14,Ko14} 
In contrast, the BI of Au/$g$-C$_{14}$N$_3$ is driven by the band exchange constrained 
by the symmetry, as mentioned above. 

In summary, by showing the existence of a topological nontrivial gap of 50~meV 
surrounding the Dirac point of graphene, we demonstrate through first-principles calculations 
that a Kane-Mele TI~\cite{Kan05} could be realized by depositing Au on $g$-C$_{14}$N$_3$,
which involves substituting four C 
atoms with three N atoms in a $3\times 3$ unit cell of graphene. 
The key to understand the TI phase lies in understanding 
the electronic structures of $g$-C$_{14}$N$_3$, wherein there exists a Dirac point with 
a $+1e$ hole. Its electronic properties play an important role as follows: 
(1) Following deposition of Au with 11 valence electrons
on $g$-C$_{14}$N$_3$, the $+1e$ hole is filled and the Fermi level returns to 
the Dirac point. 
(2) To adapt to the threefold coordination of three N atoms in $g$-C$_{14}$N$_3$, 
the $s$ and $d_{xz}$/$d_{yz}$ orbitals of Au hybridize, causing the $d$ electrons 
with strong SOC to appear at the Fermi level. 
(3) An anomalous BI between the $\pi$* band and the Au+N band with MAC is formed,
leading to a topological nontrivial gap of 50~meV surrounding the Dirac point.
We note that 
the occurrence of a conventional BI depends on a profile of the interaction between 
the two involved bands, however, this anomalous BI is protected by symmetry, 
independent of the interaction between the two involved bands.

This work was supported by NFSC (No.61274097) and NBRPC (No. 2011CB921803 and 2015CB921401).

\end{document}